\documentclass[11pt,a4paper]{article}
\usepackage{latexsym}
\usepackage{graphicx}

\begin{document}

\title{Frequency-domain study of $\alpha$-relaxation in the
	  Random Orthogonal Model} 
\author{Francesco Rao$^\dag$
  \and Andrea Crisanti$^\ddag$
  \and Felix Ritort$^\#$\\ \\
  $\dag$Department of Biochemistry, University of Zurich, \\
  Winterthurerstrasse 190, CH-8057 Zurich,\\
  Switzerland\\ \\
  $\ddag$Dipartimento di Fisica, Universit\`a di Roma,\\
  ``La Sapienza'' and INFM unit\`a di Roma I and SMC,\\
  P. le A. Moro 2, 00186, Rome, Italy \\ \\
  $\#$Departament de F\'{\i}sica Fonamental, \\
  Facultat de F\'{\i}sica, Universitat de Barcelona, Diagonal 647,  \\
  08028 Barcelona,  Spain.
}

\date{printout: 15.01.2003}
 
\maketitle 
        
\begin{abstract}
  The time-dependent susceptibility for the
  finite-size mean-field Random Orthogonal model (ROM) is studied numerically
  for temperatures above the mode-coupling temperature.
  The results show that 
  the imaginary part of the susceptibility $\chi''(\nu)$ obeys the scaling 
  form proposed for glass-forming liquids with 
  the peak frequency decreasesing as the temperature is lowered
  consistently with the 
  Vogel-Fulcher law with a critical temperature remarkably close to
  the known critical temperature $T_c$ of the model where the configurational
  entropy vanishes. \\ 

  PACS: 64.40.-i, 64.60.Cn, 75.10.Nr
\end{abstract}


The spectral properties of the primary or
$\alpha$-relaxation in supercooled liquids has been largely studied by
means of dielectric spectroscopy
(Dixon {\it et al.} 1990, Chamberlin 1991, Dixon {\it et al.} 1991, 
Sch\"onhals {\it et al.} 1991, Menon and Nagel 1993, 
Sch\"onhals {\it et al.} 1993, Kuldlik {\it et al.} 1995, 
Leheny {\it et al.} 1996, Leheny and Nagel 1997)
finding that the data for the imaginary part $\epsilon''(\nu)$ of
the dielectric susceptibility $\epsilon(\nu)$ at different temperatures and
for several glass-forming liquids can be collapsed onto a master
curve using a three-parameter scaling function.
The master plot
is able to reproduce the $\epsilon''(\nu)$ data around the relaxation
peak $\nu_p$ and also at higher frequencies. The bad collapse in the low
frequency part has been object of some debates
(Sch\"onhals {\it et al.} 1991, Menon and Nagel 93, 
Sch\"onhals {\it et al.} 1991, Kuldlik {\it et al.} 1995, 
Leheny {\it et al.} 1996),
however there is no dispute above
$\nu_p$. The frequency $\nu_p$ has a very strong temperature
dependence, commonly fitted by a Vogel-Fulcher form ${\rm
  log}_{10}(\nu_p) = {\rm log}_{10}(\nu_0) - A/(T-T_0)$, where $T_0$
is close to the Kauzmann temperature (Kauzmann 1948)
where the configurational entropy vanishes (see e.g., Angell 1988).

In this contribution we compare the frequency-domain analysis of the
finite-size Random Orthogonal Model (ROM) above the mode coupling
temperature with the above scenario. The main motivation for this
study was to make a stringent test on the ROM as a possible toy-model for
the fragile-glass scenario. 
Model Hamiltonians capable of describing  relaxation processes in supercooled 
liquids and structural glasses are difficult to obtain. However,
starting with the work of Kirkpatrick, Thirumalai and Wolynes 
(Kirkpatrick and Thirumalai 1987a, b, Kirkpatrick and Wolynes 1987) in the
late 80's, it is now clear that there is a close analogy between 
some mean-field spin-glass models and structural glasses 
(Bouchaud {\it et al.} 1998).
The basic simplification occurring in mean-field models is that 
in the limit of a very large ($N\to\infty$) number of spin 
one is left with a closed set of equations for the two-time
correlation and response functions which, above a critical temperature
$T_D$, are
equivalent to the schematic mode coupling equations introduced
by Leutheusser, G\"{o}tze and others 
(Bengtzelius {\it et al.} 1984, Leutheusser 1984, G\"otze 1991)
as a model for the ideal glass transition.

In mean-field models the barrier separating different ergodic
components diverges in the mean-field limit, hence at the critical
temperature $T_D$ a real ergodic to non-ergodic transition takes place
with diverging relaxation times. The critical temperature
$T_D$ coincides with the critical
temperature $T_{MCT}$ derived in the Mode-Coupling Theory (MCT) and in what
follows we will use only the notation $T_{MCT}$. In a real
system, however, barriers are of finite height and the glass
transition appears at $T_g < T_{MCT}$ where the typical activation time
over barriers is of the same order of the observation time.  
In these systems $T_{MCT}$ represents the temperature below which the 
dynamics is dominated by activated hopping over energy barries. Therefore
to go
beyond mean-field it is necessary to include activated processes, a
very difficult task since it implies the knowledge of excitations
involved in the dynamics.
Recent studies have shown that activated processes in mean-field models 
could be included just keeping $N$ finite 
(Crisanti and Ritort 2000a, b), giving
support to the scenario of the fragile glass transition developed
from spin-glass models. This is not
a trivial assumption since it is not {\it a priori} clear why
excitations in mean-field spin glass models should have similar properties 
to those of supercooled liquids. This, for example, seems to be the case
for the Random Orthogonal model (ROM) 
(Marinari {\it et al.} 1994), but not 
for the mean-field Potts-Glass model 
(Bragian {\it et al.} 2001, Bragian {\it et al.} 2002).

To compare the results from the finite-size ROM with the experimental one
(Dixon {\it et al.} 1990, Chamberlin 1991, Sch\"onhals {\it et al.} 1991,
Dixon {\it et al.} 1991,  Menon and Nagel 1993, Sch\"onhals {\it et al.} 1993,
Kuldlik {\it et al.} 1995, Menon and Nagel 1995, Leheny {\it et al.} 1996, 
Leheny and Nagel 1997)
in this contribution we shall consider only temperatures above $T_{MCT}$.
Since the range of temperatures we explore 
are all above the mode-coupling transition $T_{MCT}$,
we do not expect to find diverging timescales in the large
$N$ limit. 

The ROM (Marinari {\it et al.} 1994) is defined by the Hamiltonian
\begin{equation}
\label{eq:ham}
  H = - 2 \sum_{ij} J_{ij}\, \sigma_i\, \sigma_j 
      - h\sum_{i}\, \sigma_i
\end{equation}
where $\sigma_i=\pm 1$ are $N$ Ising spin variables, and $J_{ij}$ is a
$N\times N$ random symmetric orthogonal matrix with $J_{ii}=0$.  
For $N\to\infty$ and $h=0$ this model has a dynamical transition at
$T_{MCT}=0.536$, and a static transition at $T_c=0.256...$ 
(Marinari {\it et al.} 1994).
Numerical simulations are performed using the Monte Carlo (MC) method with
the Glauber algorithm for temperatures in the range $0.6$ up to $2.0$.
To study the frequency response we considered a time-dependent field of
the form $h(t) = h_0\cos(2\,\pi\,\nu t)$, where the time is measured
in MC steps and $h_0=0.2$ small enough to be within
the linear response regime. In our simulations the typical range of
$\nu$ was $10^{-6}$ -- $10^{-1}$. For each frequency $\nu$ 
the complex susceptibility $\chi(\nu) = \chi'(\nu) + i\chi''(\nu)$ 
is given by
\begin{equation}
\label{eq:chip}
\chi'(\nu) = \frac{1}{NM} \sum_{t=1}^{M}\sum_{j=1}^{N} \sigma_j(t)\, 
                               \cos(2\,\pi\,\nu t),
\end{equation}
\begin{equation}
\label{eq:chipp}
\chi''(\nu) = \frac{1}{NM} \sum_{t=1}^{M}\sum_{j=1}^{N} \sigma_j(t)\, 
                               \sin(2\,\pi\,\nu t).
\end{equation}
The number of MC steps $M$ after equilibration was $100$ for the
largest $\nu$ and up to $10^7$ for the shortest $\nu$. As system size 
we used $N=300$ which is a good compromise between small
sample-to-sample fluctuations and small barriers height.

\begin{figure}[hbt!]
  \centering
  \includegraphics[scale=0.9]{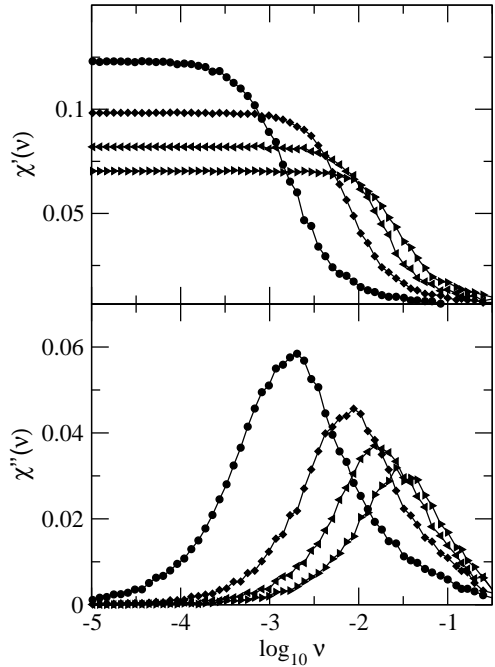}
  \caption{The real and imaginary part of the complex susceptibility 
    $\chi'$ and $\chi''$ as function of $\nu$ for the ROM with $N=300$ at
    different temperatures. Upper panel top to bottom, lower panel
    left to right, $T=0.7$, $0.9$, $1.1$ and $1.3$.
  }
  \label{fig:suc300}
\end{figure}
Figure  \ref{fig:suc300} shows the real and imaginary parts of the 
susceptibility over the available range of frequency. Not all temperatures
are reported for a better drawing.
The relaxation peak in the imaginary part can be fitted with a log-normal
form (Wu and Nagel 1992):
\begin{equation}
 \label{eq:logn}
 \chi''(\nu) = \frac{\Delta\chi}{\sqrt{\pi}\Sigma}
          \exp\left[-({\rm log}_{10}\nu - {\rm log}_{10}\nu_p)^2/\Sigma^2
               \right]
\end{equation}
where $\nu_p$ is the frequency of the peak, $\Sigma$ the width,
and $\Delta\chi = \chi'_0 - \chi'_{\infty}$, where $\chi'_0$ and
$\chi'_\infty$ are, respectively, 
the low and high frequency limit of $\chi'(\nu)$.

\begin{figure}[hbt!]
  \centering
  \includegraphics[scale=0.9]{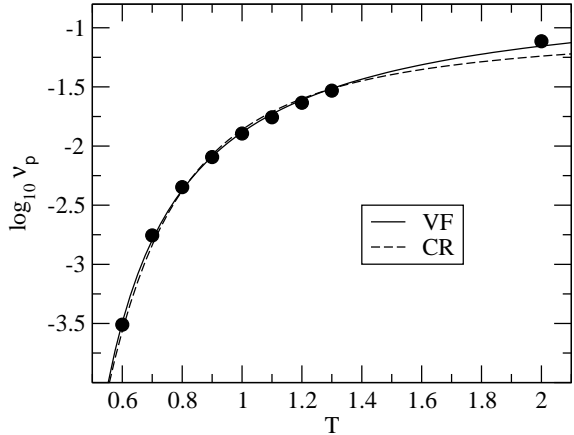}
  \caption{${\rm log}_{10}\nu_p$ as function of $T$ for the ROM with
  $N=300$. 
   The full line is the Vogel-Fulcher
    law $\nu_p = \nu_0 \exp[ -A /(T-T_0) ]$ while
    the dashed line is the formula
  $\nu_p = \nu_0 \exp[ -A \beta_{\rm eff}(T)/T ]$ of Ref. 
  (Crisanti and Ritort 2002) with 
  $\beta_{\rm eff}(T) = \partial S_c(e_{\rm is}) / \partial e_{\rm is} 
                     |_{e_{\rm is}=e_{\rm is}(T)}$ evaluated using the
 results of Refs. (Crisanti and Ritort 2000a, b).
 The discrepancy at high
 temperature is probably due to a poor numerical estimation of the 
 configurational entropy, indeed a similar deviation is found using the
 Adam-Gibbs formula (not reported). 
  }
  \label{fig:vogel-fit300}
\end{figure}

As the temperature is lowered the peak 
frequency $\nu_p$ decreases, and the width $\Sigma$ broadens.  
The behavior of $\nu_p$ is consistent with the Vogel-Fulcher law
$\exp[-A/(T-T_0)]$ (Menon and Nagel 1995).
The fit of the frequency peak $\nu_p$ for the ROM 
with the Vogel-Fulcher formula is rather good,
see Fig. \ref{fig:vogel-fit300},
and gives
$A = 0.89\pm 0.06$ and $\ln\nu_0 = 0.64\pm 0.02$
$T_0= 0.28\pm 0.02$, a value in agreement with 
the critical value $T_c = 0.256...$. We note, however, that data can also be
fitted using different expressions such as $\exp[-A/T^2]$ or the 
Adam-Gibbs formula $\exp[-A/(TS_{c}(T))]$ where $S_{c}$ is the
configurational entropy, and in particular with the
formula (CR)
$\nu_p = \nu_0 \exp[ -A \beta_{\rm eff}(T)/T ]$ 
where 
$\beta_{\rm eff}(T) = \partial S_c(e_{\rm is}) / \partial e_{\rm is} 
                     |_{e_{\rm is}=e_{\rm is}(T)}$ 
derived from a cooperative scenario of relaxation
(Crisanti and Ritort 2002), see Fig. \ref{fig:vogel-fit300}. 
The CR formula predicts
a crossover form fragile to strong behaviors as the temperature is lowered,
however, 
differences among all these expressions can be 
appreciated only for very low values of $\nu_p$ which are out of
our measurements range. 

\begin{figure}[hbt!]
  \centering
  \includegraphics[scale=0.85]{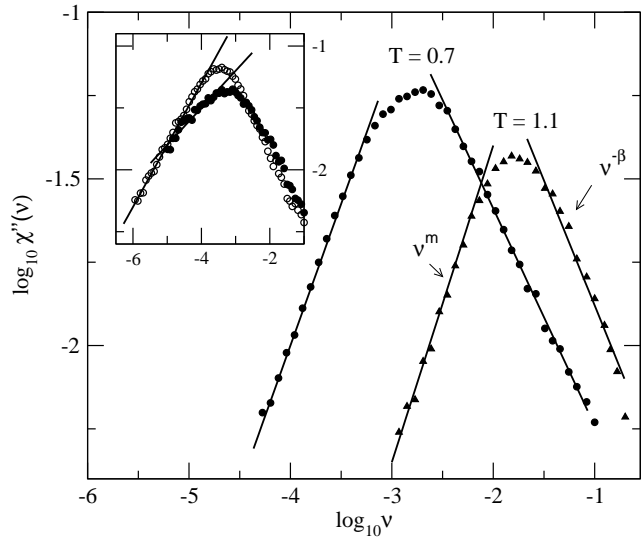}
  \caption{$\chi''(\nu)$ for the ROM with $N=300$ for
    temperatures $T=0.7$ and $1.1$. The exponents are $m=1$, $\beta=1.$ for 
    $T=1.1$ and $m=.96$, $\beta=.71$ for $T=0.7$.
    Inset: $\chi''(\nu)$ at temperature $T=0.6$ for $N=64$ (filled circles) 
    and $300$ (empty circles). The lines have slope $m=0.3$ for $N=64$
    and $m=0.5$ for $N=300$. The increase of $m$ toward $1$ as
    $N$ grows is clearly seen.
  }
\label{fig:betafit300}
\end{figure}

The analysis of the response for glass-former liquids reveals three power laws
for $\chi''$ (Leheny and Nagel 1997):
 \begin{equation}
\chi''(\nu)\sim \left\{\begin{array}{cc}
 \nu^{m}      & \nu < \nu_p\\
 \nu^{-\beta} & \nu > \nu_p\\
 \nu^{-\sigma} & \nu\gg \nu_p\\
\end{array}
\right.
\end{equation}
The discrete nature of Monte Carlo dynamics time step prevents us from 
resolving the last, 
nevertheless the first two regimes are clearly seen, as shown in Figure
\ref{fig:betafit300}. At higher temperatures $m=\beta=1$ and the
relaxation is Debye-like with exponential decaying correlations. 
As the temperature is lowered the value of $\beta$ 
decreases below $1$ and decay becomes stretched-exponential.
%
%
It is known that for glass-former liquids  $\beta$ and $\sigma$  
are related by $(\sigma+1)/(\beta+1) = \gamma$, where $\gamma$ is a
constant (Leheny and Nagel 1997).
Furthermore $\sigma$ varies linearly with temperature:
$\sigma = B(T-T_\sigma)$ with $T_\sigma\simeq T_0$ 
(Leheny and Nagel 1997),
This implies that 
$\beta = B'(T-T_0) + (1-\gamma)/\gamma$.
Inserting into this formula the values of $\beta$ obtained for the ROM 
at various temperatures and the value of $T_0$ computed from $\nu_p$ we find
$\gamma = 0.72\pm 0.02$ the same value 
found for real liquids (Menon and Nagel 1995, Leheny and Nagel 1997).

The analyticity of $\chi(\nu)$ and linearity of absorption at 
asymptotically low frequencies implies that $\chi''(\nu) \propto \nu$ for
$\nu\ll\nu_p$ (Sch\"onhals {\it et al.} 1991, Menon and Nagel 1993, 
Sch\"onhals {\it et al.} 1993).
For the ROM with $N=300$ we find
$m\simeq 1$ for temperatures down to about $T=0.8$ while below 
significant deviations with $m<1$ are observed.  
Similar deviations have been observed in data from glass-forming liquids and
generated some controversy 
(Dixon {\it et al.} 1990, Chamberlin 1991, Sch\"onhals {\it et al.} 1991,
Dixon {\it et al.} 1991,  Menon and Nagel 1993, Sch\"onhals {\it et al.} 1993,
Kuldlik {\it et al.} 1995, Leheny {\it et al.} 1996) 
on the reliability of the scaling form proposed by Dixon {\it et al.} 
(Dixon {\it et al.} 1990, Chamberlin 1991, Dixon {\it et al.} 1991). 
Many liquids posses secondary relaxations 
which overlap
the primary response broadening the peak and leading to deviation from 
linearity (Kuldlik {\it et al.} 1995, Leheny {\it et al.} 1996).
In the case of ROM these secondary relaxations
are related to the fact that the barriers separating 
the low states sampled as the temperature
is decreased toward $T_{MCT}$ are not well separated for not too large $N$.
Indeed studies of mean-field spin-glass models for the structural glass 
transition shows that in the thermodynamic limit 
there is no gap between saddles separating 
local minima with energy above the threshold energy associated with 
the dynamical transition (Cavagna {\it et al.} 1997).
This is a situation more reminiscent of spin-glasses rather than glasses
for which both experimental 
(Bitko {\it et al.} 1996) and numerical 
simulations (Bitko {\it et al.} 1996, Rao 2001) show a broader shape
of $\chi''(\nu)$ near the peak.

This scenario is supported by a finite-size scaling analysis of the ROM.
Indeed we find that for a fixed temperature 
while $\beta$ is independent on $N$, the
value of $m$, when less than $1$, increases toward $1$ as $N$ is increased.
In the inset of 
Figure \ref{fig:betafit300} we show $\chi''(\nu)$ at $T=0.6$ for 
systems of size $N=64$ and $N=300$. The increase of $m$ going from $64$ to 
$300$ spins is clear.
%
%
Finally we address the goodness of the Nagel scaling.
Dixon {\it et al} 
(Dixon {\it et al.} 1990, Chamberlin 1991, Dixon {\it et al.} 1991) 
have shown that all data for the dielectric 
susceptibility $\epsilon(\nu)$ of different glass-forming liquids and 
temperatures
can be collapsed onto a single master curve by plotting
$(1/w){\rm log}_{10}(\epsilon''(\nu)\nu_p/\nu\Delta\epsilon)$ versus
$(1/w)(1/w+1){\rm log}_{10}(\nu/\nu_p)$, where 
$\Delta\epsilon= \epsilon'(0)-\epsilon'(\infty)$,
$\nu_p$ is the peak frequency  and $w$ is the 
half-maximum width of the $\epsilon''(\nu)$ peak normalized to the 
corresponding width of the Debye peak: $w_{D} \simeq 1.14$ decades. 

\begin{figure}[hbt!]
  \centering
  \includegraphics[scale=0.9]{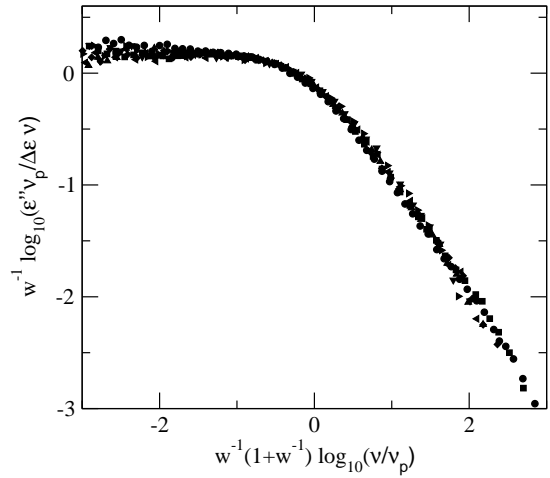}
\caption{The Nagel's scaling for the ROM with $N=300$ and temperatures
  $T=2.0$, $1.3$, $1.2$, $1.1$, $1.0$, $0.9$, $0.8$, $0.7$ and $0.6$.
        }
\label{fig:scaling300}
\end{figure}

Figure \ref{fig:scaling300} shows the Nagel plot for the ROM with $N=300$. 
The data are from temperatures ranging from $0.6$ up to $2.0$.  For each 
curve the parameters
$\Delta\chi$, $w = 2\sqrt{2}\Sigma/w_d$ and $\nu_p$ have been obtained
from the log-normal fit of $\chi''$ using (\ref{eq:logn}).
We see that while the collapse  for $\nu>\nu_p$ is good for all temperatures,
for $\nu<\nu_p$ only data with $m=1$ do collapse.
As noted in Refs. 
(Kuldlik {\it et al.} 1995, Leheny {\it et al.} 1996) the optimization of the
three parameters through the fitting is essential to have a good collapse
of data. 

In conclusion, we have shown that the primary relaxation in the finite
size mean-field ROM obeys the scaling form typical of glass-forming
liquids. Furthermore, the frequency peak of the imaginary part of the
complex susceptibility follows the Vogel-Fulcher law with critical
temperature $T_0=0.28\pm 0.02$ very close to the critical temperature
$T_c=0.256...$, the Kauzmann temperature of the model. All system sizes
studied (up to $N=300$) lead to this value for $T_0$. Because we used
Monte Carlo dynamics there is a maximum value for the frequency $\sim
N$ determined by the discreteness of the elementary time step. As a
consequence we are not able to resolve the second scaling behaviors of
the imaginary part of the susceptibility $\chi''(\nu)\sim
\nu^{-\sigma}$ for $\nu\gg\nu_p$.  Nevertheless by assuming that the
exponent $\sigma$ vanishes linearly at $T_0$, we obtained for the
constant $\gamma$ relating the exponents $\beta$ and $\sigma$ the
same value $\gamma=0.72\pm 0.02$, found for real glass-forming
liquids (Menon and Nagel 1995, Leheny and Nagel 1997).
This value does not depend on
the system size for the sizes we studied.  Overall, the present
results show that finite-size mean-field spin glasses capture the
cooperative effects responsible for the relaxational processes
observed in glass forming liquids when approaching the mode-coupling
temperature from above.  The extension of this analysis to
the region below $T_{MCT}$ where strong finite $N$ effects are to be
observed remains an interesting open problem.

\bigskip
\noindent
We acknowledge 
F. Sciortino and P.Tartaglia for a critical reading of the manuscript.
A.C. acknowledges support from the INFM-SMC center. F.R has been
supported by project the Spanish Ministerio de Ciencia y Tecnolog\'{\i}a
Grant number BFM2001-3525 and Generalitat de Catalunya. A.C and F.R have
also benefited from the Acciones Integradas Espa\~na-Italia HI2000-0087.

\section*{References}

\begin{description}

\item        Angell,  C. A., 1988
        J. Phys. Chem. Solids, {\bf 49}, 863.

\item        Bengtzelius, U., G\"otze, W., and Sjolander, A., 1984,
        J. Phys. C, {\bf 17}, 5915.

 \item       Bitko, D., Menon, N., Nagel, S. R., Rosenbaum, T. F., and Aeppli, G., 
	1996
        Europhys. Lett, {\bf 33}, 489.

\item  Bouchaud, J. P., Cugliandolo, L. F., Kurchan, K., and Mezard, M., 1988,
              {\it Spin Glasses and Random Fields}, edited by A. P. Young 
              (World Scientific, Singapore, 1998), p. 161.

\item        Brangian, C.,  Kob, W.,  Binder, K., 2001
        Europhys. Lett., {\bf 53}, 756.

\item        Brangian, C.,  Kob, W.,  Binder, K., 2002
        J.  Phys. A, {\bf 35}, 191.

\item        Cavagna, A., Giardina, I.,  and Parisi, G., 1997,
        J. Phys. A, {\bf 30}, 7021.

\item         Chamberlin, R. V., 1991, 
         Phys. Rev. Lett., {\bf 66}, 959.

\item        Crisanti, A.,  and Ritort, F., 2000a,
        Europhys. Lett., {\bf 51}, 147.

\item        Crisanti, A.,  and Ritort, F., 2000b,
        Europhys. Lett., {\bf 52}, 640.

 \item      Crisanti, A., and Ritort, F., 2002
       Philosophical Magazine B {\bf 82}, 143.

\item         Dixon, P. K., Wu, L.,  Nagel, S. R., Williams, B D.,
         and Carini, J. P., 1990,
         Phys. Rev. Lett., {\bf 65}, 1108.

\item         Dixon, P. K., Wu, L.,  Nagel, S. R., Williams, B D.,
         and Carini, J. P., 1991,
         Phys. Rev. Lett., {\bf 66}, 960.

\item        G\"{o}tze, W., 1991,
        {\it  Liquids, Freezing and Glass Transition}, edited by
        J. P. Hansen, D. Levesque and J. Zinn-Justin,
        (North-Holland, 1991), p. 287.

\item        Kauzmann, W., 1948,
        Chem Phys. Rev., {\bf 43}, 219.

 \item       Kirkpatrick, T. R., and Thirumalai, D., 1987a,
        Phys. Rev. Lett., {\bf 58}, 2091.

 \item       Kirkpatrick, T. R., and Thirumalai, D., 1987b,
        Phys. Rev. B, {\bf 36}, 5388.

\item        Kirkpatrick, T. R., and Wolynes, P. G., 1987,
        Phys. Rev. B, {\bf 36}, 8552.

\item        Kudlik, A., Benkhof, S., Lenk, R., and R\"{o}ssler, E., 1995
         Europhys. Lett., {\bf 32}, 511.

\item         Leheny, R. L., Menon, N.,  and Nagel, S. R., 1996
         Europhys. Lett., {\bf 36}, 473.

\item        Leheny, R. L., and Nagel, S. R., 1997
        Europhys. Lett., {\bf 39}, 447.

\item        Leutheusser, E., 1984,
        Phys. Rev. A, {\bf 29}, 2765.

 \item       Marinari, E., Parisi, G., and Ritort, F., 1994,
	J. Phys. A, {\bf 27}, 7647.

\item        Menon, N.,  and Nagel, S. R., 1993
        Phys. Rev. Lett., {\bf 71}, 4095.

\item         Menon, N.,  and Nagel, S. R., 1995
         Phys. Rev. Lett. {\bf 74}, 1230.

\item        F. Rao, F, 2001,
        Master Thesis, University of Rome ``La Sapienza''.

\item        Sch\"{o}nhals, A., Kremer, F.,  and Schlosser, E., 1991
        Phys. Rev. Lett.,  {\bf 67}, 999.

\item        Sch\"{o}nhals, A., Kremer, F.,  and Schlosser, E., 1993
        Phys. Rev. Lett., {\bf 71}, 4096.

\item        Wu, L. and Nagel, S. R., 1992,
        Phys. Rev. B, {\bf 46}, 11198.

\end{description}
\end{document}